\documentclass[10pt]{article}
\DeclareSymbolFont{AMSb}{U}{msb}{m}{n}
\DeclareMathSymbol{\N}{\mathbin}{AMSb}{"4E}
\DeclareMathSymbol{\Z}{\mathbin}{AMSb}{"5A}
\DeclareMathSymbol{\R}{\mathbin}{AMSb}{"52}
\DeclareMathSymbol{\Q}{\mathbin}{AMSb}{"51}
\DeclareMathSymbol{\I}{\mathbin}{AMSb}{"49}
\DeclareMathSymbol{\C}{\mathbin}{AMSb}{"43}

\usepackage{times,psfrag,graphicx,color,authblk}
\usepackage{amsfonts,amsmath,amssymb} %{amsthm}
\usepackage{psfrag,graphics,epsfig}

\makeatother
\author{Akshay Kashyap}
\affil{\small{Dept. of ECE, UIUC, Urbana, IL. Tel: 217-766-2537, Email: {\tt{kashyap@uiuc.edu}}.}}
\author{Luis Alfonso Lastras-Monta\~{n}o}
\affil{\small{IBM T.J. Watson Research Center, Yorktown Heights, NY. Email: \tt{lastrasl@us.ibm.com}.}}
\author{Cathy Xia}
\author{Zhen Liu}
\affil{\small{IBM T.J. Watson Research Center, Hawthorne, NY. Emails: \tt{cathyx@us.ibm.com}},
\tt{zhenl@us.ibm.com}.}
\begin{document}

\newcommand{\degrees}[1]{\ensuremath{{#1}^\circ}}
\newcommand{\comment}[1]{}

\newcommand{\figlbl}[1]{\label{fig:#1}}
\newcommand{\figref}[1]{Figure~\ref{fig:#1}}

\newcommand{\chlbl}[1]{\label{ch:#1}}
\newcommand{\chref}[1]{Chapter~\ref{ch:#1}}

\newcommand{\seclbl}[1]{\label{sec:#1}}
\newcommand{\secref}[1]{Section~\ref{sec:#1}}

\newcommand{\eqnlbl}[1]{\label{eq:#1}}
\newcommand{\eqnref}[1]{(\ref{eq:#1})}

\newcommand{\tbllbl}[1]{\label{tbl:#1}}
\newcommand{\tblref}[1]{Table~\ref{tbl:#1}}

\newcommand{\lemlbl}[1]{\label{lem:#1}}
\newcommand{\lemref}[1]{Lemma~\ref{lem:#1}}

\newcommand{\deflbl}[1]{\label{def:#1}}
\newcommand{\defref}[1]{Definition~\ref{def:#1}}

\newcommand{\thmlbl}[1]{\label{thm:#1}}
\newcommand{\thmref}[1]{Theorem~\ref{thm:#1}}

\newcommand{\alglbl}[1]{\label{alg:#1}}
\newcommand{\algref}[1]{Algorithm~\ref{alg:#1}}

\newcommand{\proplbl}[1]{\label{prop:#1}}
\newcommand{\propref}[1]{Proposition~\ref{prop:#1}}

\newcommand{\applbl}[1]{\label{apdx:#1}}
\newcommand{\appref}[1]{Appendix~\ref{apdx:#1}}
\newcommand{\seeapp}[1]{(see \appref{#1})}

%%%%%%%%%%%%%%% OTHER NEWCOMMANDS THAT I HAVE GOTTEN USED TO %%%%%%%%%%%%%%%%%%
\newcommand{\contime}{T_{\mathrm{con}}}
\newcommand{\xavg}{\bar{x}_{\infty}}
\newcommand{\dmx}{\d_{max}}
\newcommand{\isdef}{\triangleq}
\newcommand{\omo}{{\omega}_o}
\newcommand{\epq}{{\epsilon}_Q}
\newcommand{\epl}{\epsilon}
\newcommand{\Omg}{\Omega}
\newcommand{\ul}{\underline}
\newcommand{\noi}{\noindent}
\newcommand{\lb} {\label}
\newcommand{\al}{\alpha}
\newcommand{\nno}{\nonumber}
\newcommand{\impls}{\Rightarrow}
\newcommand{\bq}{\begin{equation}}
\newcommand{\eq}{\end{equation}}
\newcommand{\bqa}{\begin{eqnarray}}
\newcommand{\eqa}{\end{eqnarray}}
\newcommand{\bqas}{\begin{eqnarray*}}
\newcommand{\eqas}{\end{eqnarray*}}
\newcommand{\bm}{\boldmath}
\newcommand{\ubm}{\unboldmath}
\newcommand{\cjgt}[1]{#1^{\dagger}}
\newcommand{\cplx}[1]{\C^{#1}}
\newcommand{\real}[1]{\R^{#1}}
\newcommand{\bl}[1]{\mathbf{#1}}
\newcommand{\tr}{\mbox{${\mathrm{tr}}$}}
\newcommand{\sinc}{\mbox{${\mathrm{sinc}}$}}
\newcommand{\samedis}{\stackrel{d}{=}}
\newcommand{\rank}{\mbox{${\mathrm{rank}}$}}
\newcommand{\diag}{\mbox{${\mathrm{diag}}$}}
\newcommand{\prob}{\mbox{${\mathrm{Pr}}$}}
\newcommand{\chull}{\mbox{${\mathrm{Co}}$}}
\def\argmin{\mathop{\rm arg\,min}}
\newcommand{\chain}{\leftrightarrow}
\newcommand{\dgr}[1]{{#1}^{\dagger}}
\newcommand{\bh}[1]{\widehat{\bl{#1}}}
\newcommand{\wh}[1]{\widehat{#1}}
\newcommand{\dg}{\dagger}
\newcommand{\pE}{\mbox{$E$}}
\newcommand{\dummy}{\mbox{$~$}}
\newcommand{\dummier}{\mbox{$\begin{array}{c}
                                                \dummy \\
                                                \dummy \\
                                                \end{array}$}}
\newcommand{\Eh}{\mbox{\boldmath${\pE}_H$}}
\newcommand{\cA}{{\cal A}}
\newcommand{\cC}{{\cal C}}
\newcommand{\cE}{{\cal E}}
\newcommand{\cF}{{\cal F}}
\newcommand{\cG}{{\cal G}}
\newcommand{\cH}{{\cal H}}
\newcommand{\cI}{{\cal I}}
\newcommand{\cL}{{\cal L}}
\newcommand{\cM}{{\cal M}}
\newcommand{\cN}{{\cal N}}
\newcommand{\cP}{{\cal P}}
\newcommand{\cQ}{{\cal Q}}
\newcommand{\cR}{{\cal R}}
\newcommand{\cS}{{\cal S}}
\newcommand{\cT}{{\cal T}}
\newcommand{\cU}{{\cal U}}
\newcommand{\cW}{{\cal W}}
\newcommand{\cX}{{\cal X}}
\newcommand{\cY}{{\cal Y}}
\newcommand{\cZ}{{\cal Z}}
\newcommand{\cCN}{{\cal CN}}

%specific to detection extreme points paper
\newcommand{\pcset}{\bar{\cQ}\cap\cH}
\newcommand{\pcsetc}{\bar{\cQ}\cap\cH^c}
\newcommand{\qset}{\bar{\cQ}}
\newcommand{\extpt}{q(\gamma^*)}

\newcommand{\nprime}{\lfloor\theta N\rfloor}
\newcommand{\xsupi}{X^{(i)}}
\newcommand{\Sx}{\bl {{\Sigma}_x}}
\newcommand{\Sxi}{\bl {{\Sigma}_x^{-1}}}
\newcommand{\Sw}{\bl {{\Sigma}_w}}
\newcommand{\Swi}{\bl {{\Sigma}_w^{-1}}}
\newcommand{\Sn}{\bl {{\Sigma}_n}}
\newcommand{\Sni}{\bl {{\Sigma}_n^{-1}}}
\newcommand{\Sy}{\bl {{\Sigma}_y}}
\newcommand{\Se}{\bl {{\Sigma}_e}}
\newcommand{\se}{{{\sigma}_e^2}}
\newcommand{\Seh}{\bl {{\Sigma}_{e|H}}}
\newcommand{\seh}{{{\sigma}_{e|H}^2}}
\newcommand{\Swy}{\bl {{\Sigma}_{wy}}}
\newcommand{\Syw}{\bl {{\Sigma}_{yw}}}
\newcommand{\Sxy}{\bl {{\Sigma}_{xy}}}
\newcommand{\sxy}{{\sigma}_{xy}}
\newcommand{\Syi}{\bl {{\Sigma}_{y}^{-1}}}
\newcommand{\Sr}{\bl {{\Sigma}_{r}}}
\newcommand{\sr}{{\sigma}_r^2}
\newcommand{\Sz}{\bl {{\Sigma}_{z}}}
\newcommand{\sz}{{\sigma}_z^2}
\newcommand{\Syx}{\bl {{\Sigma}_{yx}}}
\newcommand{\syx}{{{\sigma}_{yx}}}
\newcommand{\dimH}{r\times t}
\newcommand{\dimA}{t\times l}
\newcommand{\dagH}{\dgr{H}}
\newcommand{\dagA}{\dgr{A}}
\newcommand{\dagx}{\dgr{x}}
\newcommand{\dbgH}{\dgr{\bl{H}}}
\newcommand{\dbgx}{\dgr{\bl{x}}}
\newcommand{\blH}{\bl{H}}
\newcommand{\hf}{H_f}
\newcommand{\blx}{\bl{x}}
\newcommand{\bls}{\bl{s}}
\newcommand{\blv}{\bl{v}}
\newcommand{\blu}{\bl{u}}
\newcommand{\blw}{\bl{w}}
\newcommand{\blW}{\bl{W}}
\newcommand{\bly}{\bl{y}}
\newcommand{\bln}{\bl{n}}
\newcommand{\blz}{\bl{z}}
\newcommand{\ble}{\bl{e}}
\newcommand{\blr}{\bl{r}}
\newcommand{\blone}{\bl{1}}
\newcommand{\Scap}{\bl {\widehat{{\Sigma}}_{w}}}
\newcommand{\comchan}{\dgr{H}\Sni H}
\newcommand{\chanlam}[1]{\lambda_{#1}}
\newcommand{\srclam}[1]{\lambda_{#1}^{(s)}}
\newcommand{\ksp}{k_{sp}}
\newcommand{\ksc}{k_{sc}}
\newcommand{\ricg}{\bl{H}_r}
\newcommand{\ricgc}{\bl{H}_r^{\dagger}}
\newcommand{\ricgs}{\ksc \bl{H} + \ksp S}
\newcommand{\ricgsc}{\ksc \dgr{\bl{H}} + \ksp \dgr{S}}
\newcommand{\rigs}{\ksc \bl{H} + \ksp I}
\newcommand{\rigsc}{\ksc \dgr{\bl{H}} + \ksp I}
\newcommand{\cra}{C_{ra}}
\newcommand{\cri}{C_{ri}}
\newcommand{\cf}{C_f}
\newcommand{\lbd}{\lambda}
\newcommand{\Lbd}{\Lambda}
\newcommand{\Ud}{\dgr{U}}
\newcommand{\Vd}{\dgr{V}}
\newcommand{\srcL}{\Lambda_s}
\newcommand{\chanL}{\Lambda_c}
\newcommand{\dlin}{D_{lin}}
\newcommand{\dlinf}{D_{lin,f}}
\newcommand{\dlinra}{D_{lin,ra}}
\newcommand{\dlinri}{D_{lin,ri}}
\newcommand{\swq}{\sigma_w^2}
\newcommand{\alpq}{\alpha^2}
\newcommand{\sxq}{\sigma_x^2}
\newcommand{\syq}{\sigma_y^2}
\newcommand{\srq}{\sigma_r^2}
\newcommand{\szq}{\sigma_z^2}
\newcommand{\snq}{\sigma_n^2}
\newcommand{\bodg}{B^{\dagger}_1}
\newcommand{\bo}{B_1}
\newcommand{\bolbbo}{\bodg\tilde{\Lbd}_c\bo}
\newcommand{\lsi}{\Lbd_s^{-1}}
\newcommand{\ben}{\begin{enumerate}}
\newcommand{\een}{\end{enumerate}}

\newcommand{\domH}{\C^{r\times t}}
\newcommand{\xdis}{\blx\sim{\cal{CN}}(0,Q)}
\newcommand{\xdss}{\blx\sim{\cal{CN}}(0,P)}
\newcommand{\xsdis}{\blx\sim{\cal{CN}}(0,\frac{P}{t}I_t)}
\newcommand{\xcov}{\frac{P}{t}I_t}
\newcommand{\rcov}{\frac{\ej}{r}I_r}
\newcommand{\ej}{E_J}
\newcommand{\jamav}{\frac{\ej}{r}}
\newcommand{\comav}{\frac{P}{t}}
\newcommand{\mhq}{|H|^2}
\newcommand{\rpt}[1]{{\mathrm{Re}}\left(#1\right)}
\newcommand{\cpt}[1]{{\mathrm{Im}}\left(#1\right)}
\newcommand{\st}[1]{\stackrel{st}{#1}}
\newcommand{\Lbdo}{\Lbd^{(1)}}
\newcommand{\lbdo}{\lbd^{(1)}}
\newcommand{\Lbdt}{\Lbd^{(2)}}
\newcommand{\lbdt}{\lbd^{(2)}}
\newcommand{\Lbdd}{\Lbd'}
\newcommand{\lbdd}{\lbd'}
\newcommand{\mch}{\leftrightarrow}

\newtheorem{thm}{Theorem}
\newtheorem{defn}{Definition}
\newtheorem{res}{Result}
\newtheorem{lem}{Lemma}
\newtheorem{prop}{Proposition}
\newtheorem{algo}{Algorithm}

\newenvironment{definition}[1][Definition]{\begin{trivlist}
\item[\hskip \labelsep {\bfseries #1}]}{\end{trivlist}}
\newenvironment{example}[1][Example]{\begin{trivlist}
\item[\hskip \labelsep {\bfseries #1}]}{\end{trivlist}}
\newenvironment{remark}[1][Remark]{\begin{trivlist}
\item[\hskip \labelsep {\bfseries #1}]}{\end{trivlist}}
\newenvironment{corollary}[1][Corollary]{\begin{trivlist}
\item[\hskip \labelsep {\bfseries #1}]}{\end{trivlist}}

\newcommand{\qed}{\nobreak \ifvmode \relax \else
      \ifdim\lastskip<1.5em \hskip-\lastskip
      \hskip1.5em plus0em minus0.5em \fi \nobreak
      \vrule height0.75em width0.5em depth0.25em\fi}

%\twocolumn

\title{Distributed source coding in dense sensor networks}
%\author{}
\date{}
\maketitle{}
\begin{abstract}
We study the problem of the reconstruction of a Gaussian field defined in $[0,1]$ using $N$ 
sensors deployed at regular intervals. The goal is to quantify the total data rate required 
for the reconstruction of the field with a given mean square distortion. 
We consider a class of two-stage mechanisms which a) send information to allow the 
reconstruction of the sensor's samples within sufficient accuracy, and then b) use these 
reconstructions to estimate the entire field. To implement the first stage, the heavy 
correlation between the sensor samples suggests the use of distributed coding schemes to 
reduce the total rate. We demonstrate the existence of a distributed block coding scheme that 
achieves, for a given fidelity criterion for the reconstruction of the field, a total 
information rate that is bounded by a constant, independent of the number $N$ of sensors. 
%The rate of this scheme is within a constant, independent of $N$, of the minimum information 
%rate required by an encoder that has access to all the sensor measurements simultaneously. 
The constant in general depends on the autocorrelation function of the field and the desired 
distortion criterion for the sensor samples. We then describe a scheme which can be implemented 
using only scalar quantizers at the sensors, without any use of distributed source coding, 
and which also achieves a total information rate that is a constant, independent of the number of 
sensors. While this scheme operates at a rate that is greater than the rate achievable through
distributed coding and entails greater delay in reconstruction, its simplicity makes it attractive
for implementation in sensor networks.
\end{abstract}

\section{Introduction}\seclbl{intro}

In this paper, we consider a sensor network deployed for the purpose of sampling
and reconstructing a spatially varying random process. For the sake of concreteness,
let us assume that the area of interest is represented by the line segment $[0,1]$, 
and that the for each $s\in [0,1]$, the value of the random process is $X(s)$. For
example, $X(s)$ may denote the value of some environmental variable, such as temperature,
at point $s$. 

A sensor network, for the purpose of this paper, is a system of sensing devices (sensors) capable of
\ben
\item taking measurements from the environment that they are deployed in, and 
\item communicating the sensed data to a fusion center 
for processing.
\een
The task of the fusion center is to obtain a reconstruction $\{\tilde{X}(s), s\in [0,1]\}$
of the spatially varying process, while meeting some distortion criteria. 

There has been great interest recently in performing such sensing tasks with small, low
power sensing devices, deployed in large numbers in the region of 
interest~\cite{neuhoff},~\cite{servetto},~\cite{neupra}~\cite{ishwaretaljourn}. 
This interest is motivated by the 
commercial availability
of increasingly small and low-cost sensors which have a wide array of sensing
and communication functions built in (see, for example,~\cite{mica2dot}), and yet
must operate with small, difficult to replace batteries. 

Compression of the sensed data is of vital importance in a sensor network.
Sensors in a wireless sensor network operate under severe power constraints, and
communication is a power intensive operation. The rate at which sensors must 
transmit data to the fusion center in order to enable a satisfactory reconstruction 
is therefore a key quantity of interest. Further, in any communication 
scheme in which there is an upper bound (independent of the number of sensors)
on the amount of data that the fusion center can receive per unit time, there is another
obvious reason why the compressibility of sensor data is important - the average
rate that can be guaranteed between any sensor and the fusion center varies inversely
with the number of sensors. Therefore, any scheme in which the per-sensor rate decreases
slower than inversely with the number of sensors will build backlogs of data at sensors
for large enough number of sensors. 

Environmental variables typically vary slowly as a function of space and it is reasonable 
to assume that samples at locations close to each other will be highly correlated. 
The theory of distributed source coding (\cite{slepianwolf},~\cite{zberger},~\cite{tavya}) 
shows that if the sensors have knowledge of this correlation, then it is possible 
to reduce the data-rate at which the sensors need to communicate, while still maintaining 
the property that the information conveyed by each sensor depends only on that sensor's 
measurements. Research on practical techniques 
(\cite{discus},~\cite{colemanlc},~\cite{stankovic},~\cite{litublum},~\cite{chenhejagmohan}) 
for implementing distributed source coding typically focuses on two correlated sources, 
with good solutions for the many sources problem still to be developed. 
Thus, in our work, 
we attack the problem at hand using the available theoretical tools which have their origins 
in~\cite{slepianwolf}. 

This approach has been taken earlier in~\cite{neuhoff} and~\cite{servetto}, which investigate
whether it is possible to use such distributed coding schemes to reduce the {\em per-sensor} 
data rate by  deploying a large {\em number} of sensors at closely spaced locations in the area of interest. 
In particular, it is investigated whether it is possible to construct coding schemes in 
which the per-sensor rate decreases inversely with the number of sensors. 
The conclusion of~\cite{neuhoff}, however, is that if the sensors quantize the samples
using scalar quantizers, and then encode them, the sum of the data rates of all sensors
increases as the number of sensors increases (even with distributed coding), and therefore
the per-sensor rate cannot be traded off with the number of sensors in the manner described above.

Later, though, it was demonstrated in~\cite{mydcc05} that there exists a distributed coding
scheme 
%which indeed does achieve the desired tradeoff, in that 
which achieves a sum rate that is a constant independent of the number of sensors used (so long as there is a 
large enough number of sensors). The per-sensor rate of such a scheme therefore 
decreases inversely with the number of sensors, which is the trade-off of sensor 
number with per-sensor rate that was desired, but shown unachievable with scalar quantization,
in~\cite{neuhoff}. Results similar to those of~\cite{mydcc05} for the case when a field of
infinite size is sampled densely have since appeared in~\cite{neupra}. 
However, a question that still appears to be unresolved is whether it is possible to 
achieve a per-sensor rate that varies inversely with the number of sensors using a 
{\em simple} sensing (sampling, coding, and reconstruction) scheme. 

This paper is an 
expanded version of~\cite{mydcc05}. We describe the distributed coding scheme of~\cite{mydcc05} 
in detail, and then study another sampling and coding scheme which achieves the desired 
decrease of per-sensor rate with the number of sensors. The two main properties of this scheme
are that (1) it does not make use of distributed coding and therefore does not require
the sensors to have any knowledge of the correlation structure of the spatial variable of interest, 
and (2) it can in fact be implemented using only scalar quantizers at the sensors for the purpose
of coding the samples. The scheme utilizes the fact that the sensors are synchronized, which is 
already assumed in the models of~\cite{neuhoff},~\cite{servetto},~\cite{neupra}, and is easily 
achievable in practice. Since scalar quantizers are easily implementable in sensors with very low 
complexity, this paper shows that it is possible achieve per-sensor rates that decrease
inversely with the number of sensors with simple, practical schemes. 

A brief outline of this paper is as follows: We pose the problem formally and establish
notation in~\secref{problem}. We study the achievability of the above tradeoff with 
a distributed coding scheme in~\secref{dsc}, and compare the rate of
this coding scheme with that of a reference centralized coding scheme in~\secref{ref}. 
We describe the simple coding scheme mentioned above in~\secref{p2p}. Some numerical results
are presented in~\secref{numerical}. We make some concluding remarks in~\secref{conclusion}.

\subsection{Problem statement}\seclbl{problem}

\subsubsection{Model for the spatial process}\seclbl{assfield}
We take a discrete time model, and assume that the spatial process of interest is modeled
by a (spatially) stationary, real-valued Gaussian random process, $X^{(i)}(s)$ at each time $i$, where
$s$ is the space variable. The focus of this paper is the sampling and reconstruction of 
a finite section of the process, which we assume without loss of generality to be the interval
$[0,1]$. We follow conventional usage in referring to the spatial process 
$X^{(i)}=\{X^{(i)}(s),s\in [0,1]\}$ as the {\em field} at time $i$. 

We assume that the field $X^{(i)}$ at time $i$ is independent of the field $X^{(j)}$ for any 
$j\neq i$, and has identical statistics at all times. (In what follows, we omit the time index
when we can do so without any ambiguity.) For simplicity, we assume that $X$ is centered,
$\cE[X(s)] = 0$, and that the variance of $X(s)$ is unity, for all $s\in [0,1]$. 
The autocorrelation function of the field is denoted as 
\[
\rho(\tau) = \cE\left[ X(s) X(s + \tau)\right].
\]
Following common usage, we sometimes refer to $\rho$ as the {\em correlation structure} of the
field. 
Clearly, $\rho(0) = 1$, and $\rho(\tau) \leq 1$ for any $\tau$. 
We need only mild assumptions on the field $X$: 
\ben
\item We assume that $X$ is mean-square continuous, which is equivalent to the continuity 
of $\rho$ at $0$ (see, for example,~\cite{hajekrpnotes}). 
\item We assume that there is a neighborhood of $0$ in which $\rho$ is 
non-increasing. 
\een

Note that all results in this paper extend to fields in higher dimensions. We restrict the exposition
to one-dimensional fields for clarity and to avoid the tedious notation required for 
higher dimensional fields. 

\subsubsection{Assumptions on the sensor network}\seclbl{assnet}

We assume that $N$ sensors are placed at regular intervals in the segment $[0,1]$, with sensor 
$k$ being placed at $s_k = \frac{2k-1}{2N}$ for $k=1,2,\ldots,N$.  Sensors are assumed
to be synchronized, and at each time $i$, sensor $k$ can observe the value $X^{(i)}(s_k)$
of the field at its location, for each $k$. Sensor $k$ encodes a block of $m$ observations, 
$[X^{(1)}(s_k),X^{(2)}(s_k),\ldots,X^{(m)}(s_k)]$ into an index $I_k$ chosen from the set 
$\{1,2,\ldots,\lfloor e^{mR_k}\rfloor\}$, where $R_k$ is the rate of sensor $k$, which we 
state in the units of nats per discrete time unit. We assume that the blocklength $m$ is the 
same at all sensors. The messages of the sensors are assumed to be communicated to the fusion 
center over a shared, rate constrained, noiseless channel. The fusion center then uses the received 
data to produce a reconstruction $\tilde{X}^{(i)}(s)$ 
of the field.

A {\em coding scheme} is a specification of the sampling and encoding method used at all sensors,
as well as the reconstruction method used at the fusion center. 

\subsubsection{Error criterion}

We refer to $\cE(X^{(i)}(s) - \tilde{X}^{(i)}(s))^{2}$ as the mean square error (MSE) 
of the reconstruction of the field at point $s$ and time $i$. We measure the error
in the reconstruction as the average (over a blocklength) integrated MSE, which is 
defined as
\bq
\eqnlbl{costfn}
J_{MSE}(m) = \frac{1}{m}\sum_{i=1}^m
\int_0^1 \cE\left(\xsupi(s)-\tilde{X}^{(i)}(s)\right)^2ds.
\eq
We study coding schemes in which, for all large enough blocklengths $m$ and a specified positive
constant $D_{net}$, the fusion center is able reconstruct the field with an integrated
MSE of less than $D_{net}$, that is, schemes for which
\bq
\eqnlbl{errorcrit}
\lim_{m \rightarrow \infty}J_{MSE}(m)\leq D_{net}.
\eq

\subsubsection{Sum rate}\seclbl{sumrate}

In this paper, we describe coding schemes in which for any given value of $D_{net}$ in~\eqnref{errorcrit},
the sum rate, $\sum_{k=1}^{N} R_{k}$, is bounded above by some constant 
$\bar{R}$ independent of the number $N$ of sensors. The bound $\bar{R}$ may in general
depend on $D_{net}$.  This allows the per-sensor rate can be traded off with the number of sensors, 
so that for all $N$ large enough, the rate of each sensor is no more than a constant multiple of $\frac{1}{N}$.
%\marginpar{\color{red}Should we mention that this is {\em the} desired scaling. We repeat some of this
%in the beginning of~\secref{dsc}}

\subsection{Contributions}

Our main contributions are:
\ben
\item We prove the existence of a distributed coding scheme in which, under the assumption
that the correlation structure is known at each sensor, a sum rate that is 
independent of the number of sensors $N$ can be achieved. 
\item We design a simple coding scheme which can be implemented using scalar quantization at 
sensors, which does not require the sensors to have any information about the correlation structure, 
and which makes use of the fact that the sensors are synchronized to achieve a sum rate that is a 
constant independent of $N$. 
\een

The latter scheme has the advantage of being simple enough to be implementable even with
extremely resource-constrained sensors. However, the sum-rate achievable through this scheme is 
in general greater than the sum-rate achievable through distributed coding. Also, unlike distributed coding, 
this scheme entails a delay that increases with the number of sensors in the network. 

%\subsection{Related work}\seclbl{relwork}

\section{Distributed coding}\seclbl{dsc}

In this section we describe a distributed coding scheme which achieves the desired scaling.

\subsection{Encoding and decoding}
The scheme consists of $N$ encoders, $\{f_k\}_{k=1}^{N}$, where $f_k$ is the encoder at 
sensor $k$, and $N$ decoders, $\{g_k\}_{k=1}^{N}$ at the fusion center. For each $k$,
the rate of $f_k$ is assumed to be $R_k$, and $f_k$ maps the block 
\[
[X^{(1)}(s_k),X^{(2)}(s_k),\ldots,X^{(m)}(s_k)]
\]
of samples to an index $I_k$ chosen 
from $\{1,2,\ldots,\lfloor e^{mR_k}\rfloor\}$, which is then communicated to the fusion center.
While the output of encoder $k$ may {\em not} depend on the {\em realizations} of the 
observations at any other sensor $i\neq k$, it is assumed that all sensors have 
knowledge of the statistics of the field (in particular, the function $\rho$ is assumed 
known at each sensor\footnote{In practice, the sensors need only know the vector 
$\left[\rho\left(\frac{1}{N}\right),\rho\left(\frac{2}{N}\right),\ldots,\rho\left(\frac{N-1}{N}\right)\right]$.}) 
and utilize this information to compress their samples. The decoders
may use the messages received from all encoders to produce their reconstruction:
\begin{eqnarray}
\nno \tilde{X}^{(1,\cdots,m)}(s_{k}) = 
g_{k}(f_{1}(X^{(1,\cdots,m)}(s_{1})),\cdots,f_{{N}}(X^{(1,\cdots,m)}(s_{N}))),
%\eqnlbl{distributed}
\end{eqnarray}
where $X^{(1,\cdots,m)}(s_k)$ is shorthand for $[X^{(1)}(s_k),X^{(2)}(s_k),\ldots,X^{(m)}(s_k)]$, 
for $k=1,\ldots,N$ and similarly for $\tilde{X}$.

\subsection{Reconstructing the continuous field}\seclbl{dscrec}

The reconstruction of the field for those values of $s \in [0,1]$ where there are 
no sensors is done in a two-step fashion as follows.
In the first step, the estimates $\tilde{X}(s_k)$ of sensor samples are obtained as described above. 
Then, the value of the field between sensor locations is found by interpolation.

The interpolation $\tilde{X}(s)$ for $s\notin \{s_k | k=1,\ldots,N\}$ is based on the minimum
MSE estimator for $X(s)$ given the value of the sample closest to $s$. Formally, 
for any $s$, define $n(s) = \frac{2k+1}{2N}$ if $s\in[\frac{k}{N},\frac{k+1}{N})$ as the
location of the sample closest to $s$. Then, given $X(n(s))$, the minimum MSE estimate for
$X(s)$ is given by $\cE[X(s) | X(n(s))] = \rho(s-n(s))X(n(s))$. The reconstruction of the field at the fusion center
is obtained by replacing $X(n(s))$ in this estimate with the quantized version $\tilde{X}(n(s))$,
\bq
\eqnlbl{interpfn}
\tilde{X}(s)= \rho(s-n(s))\tilde{X}(n(s)).
\eq
While this two-step reconstruction procedure is not optimal in general, it suffices for
our purposes. 

\subsection{Error analysis}\seclbl{eadsc}
Define
\bq
\eqnlbl{jprimemse}
J_{MSE}'(m) =\frac{1}{N}\sum_{k=1}^{N}
\frac{1}{m}\sum_{i=1}^{m}\cE\left(X^{(i)}(s_k)-\tilde{X}^{(i)}(s_k)\right)^2.
\eq
Using the upper bound found in equation~\eqnref{ubf} (\appref{bounds})
on the error of the coding scheme described above, we see that 
$\lim_m J_{MSE}(m)\leq D_{net}$  is met if $\lim_mJ_{MSE}'(m)\leq D'(N)$, 
where 
%\marginpar{\color{red}NOT REQD: This error analysis uses only second order properties 
%of the field.}
\bq
\eqnlbl{dprime}
D'(N) = \left(\sqrt{D_{net}-\left(1-\rho(\frac{1}{2N})^2\right)^2} 
- \sqrt{\rho^2(\frac{1}{2N})(1-\rho^2(\frac{1}{2N}))}\right)^2,
\eq 
given that $N$ is large enough so that $1-\rho^2\left(\frac{1}{2N}\right) < D_{net}$.
It is easy to see that $D'(N)$ approaches $D_{net}$ from below as $N\rightarrow\infty$.

%for any $\epsilon$ in the interval $(0,D_{net})$, there is a
%$N_1 > 0$ such that for $N \geq N_1$, $D'(N) \geq D_{net} - \epsilon$.  

\subsection{Sum rate}

We now study the sum rate of the distributed coding scheme discussed above. We begin
with finding the encoding rates required for achieving 
\bq
\eqnlbl{discreteconstr}
\lim_mJ_{MSE}'(m)\leq D,
\eq
for some constant $D$. 

The rate region $\cR(D)$ is defined as the set of all $N-$tuples of rates 
$(R_1,R_2,\ldots,R_{N})$ for which there exist encoders
$f_k$ and decoders $g_k$, for $k=1,\ldots,N$, such that~\eqnref{discreteconstr} can be met.  
If a rate vector belongs to the rate region, we say that the corresponding set of rates 
is achievable.

The rate-distortion problem in~\eqnref{discreteconstr} is a Gaussian version of the 
Slepian-Wolf distributed coding problem~\cite{slepianwolf}. 
Until recently, the rate region for this problem was not known for even $2$ sources. 
An achievable region for two discrete sources first appeared 
in~\cite{berger77}, and was extended to continuous sources in~\cite{zberger}.
The extension to a general number of Gaussian sources appears in~\cite{pvmsc}. The two-source 
Gaussian distributed source coding problem was recently solved in~\cite{tavya}, 
where the achievable region of~\cite{berger77} was found to be tight. The rate region
is still not known for more than $2$ sources. We use the achievable region found in~\cite{pvmsc}.

Though the result is stated in~\cite{pvmsc} for individual distortion constraints on the sources, the 
extension to a more general distortion constraint is straightforward. We state the achievable 
region for distributed source coding in the form most useful to us in~\thmref{tungberg} below. 
In the statement of the theorem, we use $A \mch B \mch C$ to denote a Markov-chain relationship 
between random variables $A, B$ and $C$, that is, conditioned on $B$, $A$ is independent of 
$C$. Also, for any $S \subset \{1,\ldots,N\}$, $\bl{X}_S$ denotes the vector of those sources 
the indexes of which lie in the set $S$ and $S^c$ denotes the complement of the set $S$. 
%\marginpar{\color{red}REQD? How should Tung-Berger be stated in general? Can we make do with just stating
%second order statistics of the random variables involved? For example, can $\bl{Z}$ be {\em any} 
%random variable with mean zero and variance $pI$?}
\begin{thm}
\thmlbl{tungberg}
$\cR(D)\supset\cR_{in}(D)$, where $\cR_{in}(D)$ is the set of $N-$tuples of rates for which
there exists a vector $\bl{U} \in \R^N$ of random variables that satisfies the following conditions. 
\ben
\item $\forall~S\subseteq\{1,2,\ldots,N\}$,~~~~$\bl{U}_S \mch \bl{X}_S \mch \bl{X}_{S^c} \mch \bl{U}_{S^c}$.
\item $\forall~S\subseteq\{1,2,\ldots,N\}$,~~~~$\sum_{i\in S}R_i \geq I(\bl{X}_S;\bl{U}_S|\bl{U}_{S^c})$.
\item $\exists~~\tilde{\bl{X}}(\bl{U})$ such that 
\bq
\eqnlbl{thmdiscon}
\frac{1}{N}\sum_{i=1}^{N} \cE\left[\left(X(s_i)-\tilde{X}(s_i)(\bl{U})\right)^2\right] \leq D.
\eq
\een
\end{thm}
Note that each of the rate-constraints in~\thmref{tungberg} forms some part 
of the boundary of the achievable region $\cR_{in}$ (see, for example,~\cite{pvmsc}). In particular,
the constraint on the sum rate is not implied by any other set of constraints.

Constructing a vector $\bl{U}$ satisfying the conditions of~\thmref{tungberg} corresponds to the 
usual construction of a forward channel for proving achievability in a rate-distortion problem. 
For each $i$, $U_i$ can be thought of as the encoding of $X(s_i)$. 

We now construct a $\bl{U}$ that would suffice for our purposes. Consider a random vector $\bl{Z}\in \R^N$
that is independent of $\bl{X}$, and has a Gaussian distribution with mean $0$ and covariance matrix 
$pI$, where $I$ is the identity matrix. Then $\bl{U} = \bl{X} + \bl{Z}$ satisfies the Markov chain constraints
%\marginpar{\color{red}NOT REQD\\For non-Gaussian $\bl{X}$, this would only be the best error achievable
%by {\em linear} estimators,
%but that does not change our argument.}
of~\thmref{tungberg}. To find a good bound on the sum rate, we now find a lower bound on the 
variance $p$ for which there exists an estimator $\tilde{\bl{X}}(\bl{X} +\bl{Z})$ which satisfies
condition~\eqnref{thmdiscon}. Since $\bl{X}+\bl{Z}$ is jointly Gaussian with $\bl{X}$, the estimator
which minimizes the MSE in~\eqnref{thmdiscon} is the linear estimator, 
\bq
\eqnlbl{optest}
\tilde{\bl{X}}(\bl{X}+\bl{Z}) = \Sigma_{\bl{X}(\bl{X}+\bl{Z})}
\Sigma_{\bl{X}+\bl{Z}}^{-1}\left(\bl{X}+\bl{Z}\right),
\eq
where $\Sigma_{\bl{X}(\bl{X}+\bl{Z})} = \cE[\bl{X}(\bl{X}+\bl{Z})^T]$ and $\Sigma_{\bl{X}} = \cE[\bl{X}\bl{X}^T]$.
Let $p_{\max}(N,D,\rho)$ be the largest value of $p$ for which the MSE achieved by this estimator
satisfies~\eqnref{thmdiscon}. We prove below that for large enough $N$, $p_{\max}$ grows faster than 
linearly with $N$.

\begin{lem}
\lemlbl{lingrowth}
Let $\rho(\tau)$ be a symmetric autocorrelation function such that $\lim_{t \rightarrow 0} \rho(t) = 1$ 
and a threshold $\theta > 0$ exists for which
\begin{enumerate}
\item $1 \geq \rho(\tau) \geq \rho(\theta)>0$ if $\tau \in (0,\theta)$ and
\item the inequality $1 - \rho^{2}(\theta)/(1+\theta) \leq D$ holds.
\end{enumerate}
Then 
\begin{eqnarray*}
\liminf_{N \rightarrow \infty} \frac{1}{N}p_{\max}(N,D,\rho) \geq \theta^{2}. 
\end{eqnarray*}
\end{lem}
\emph{Note:} The second condition can be met for all $D>0$ since $1 - \rho^{2}(\theta)/(1+\theta) \rightarrow 0 $ 
as $\theta \rightarrow 0$.\\
\emph{Proof:}
We call a value of $p$ allowable if the expected reconstruction error 
in~\eqnref{thmdiscon}, with $\bl{U} = \bl{X} + \bl{Z}$, is less than $D$.
We find the largest $p$ for the error criterion: $\cE[(\tilde{X}(s_i)-X(s_i))^2] \leq D$
for each $i \in \{1,\ldots,N\}$, which is more stringent than the average error requirement
of~\eqnref{thmdiscon}.

Let us consider the estimation of $X(s_1)$. Since $\tilde{X}(s_i)$ is the best linear estimate of 
$X(s_i)$ from the data $\bl{X+Z}$, any other linear estimator cannot result in a smaller 
expected MSE. We take advantage of this observation and choose a linear estimator that although 
suboptimal, is simple to analyze and yet suffices to establish the lemma.

Our estimator for $X(s_1)$ shall be the scaled average 
$\alpha \sum_{ 1 \leq i \leq N \theta} X(s_i) + Z_{i}$,
where $\alpha$ is a parameter to be optimized shortly. To estimate $X(s_i)$ for $i \neq 0$, 
simply substitute the samples used with those whose indexes lie in the set
$\{i+1,\cdots,i+N\theta\}$ (or, for samples at the right edge of the interval $[0,1]$,
$\{i-N\theta,\cdots,i-1\}$; this does not lead to any change in what follows because of
the stationarity of the field).

%\marginpar{\color{red}NOT REQD\\All of this uses only second order properties of the random variables
%involved, and so holds irrespective of whether the field is Gaussian or not.}
It is not difficult to see that
\begin{eqnarray}
\nno \lefteqn{ \cE \left(X(s_1) - \alpha \sum_{1 \leq i \leq N \theta} X(s_{i}) + Z_{i} \right)^{2} } \\
\nno &=& \cE \left[ X(s_{1})^{2} \right] 
- 2\alpha \sum_{1 \leq i \leq N \theta} \rho(i/N) + \alpha^{2}  \cE \left( \sum_{1 \leq i \leq N \theta} X(s_{i}) \right)^{2}
+  \alpha^{2} \cE \left( \sum_{1 \leq i \leq N \theta} Z_{i} \right)^{2}   \\
\nno &\leq& 1 - 2 \alpha (N \theta-1) \rho(\theta) + \alpha^{2} N^{2} \theta^{2} + \alpha^{2} N \theta p  \\
&=& \left[ 1 - 2 \alpha N \theta \rho(\theta) + \alpha^{2} N^{2} \theta^{2} + \alpha^{2} N \theta p \right] + 2 \alpha \rho(\theta),
\eqnlbl{notdifficult}
\end{eqnarray}
where we have used the inequality $1 \geq \rho(\tau) \geq \rho(\theta)$ for $\tau \in (0,\theta)$ and the 
fact that the greatest integer not greater than $N \theta$ is at least $N \theta - 1$.
The value of $\alpha$ that makes the bracketed expression in~\eqnref{notdifficult} smallest is equal to
$\alpha^{*} = \frac{\rho(\theta)}{N \theta + p}$ (we do not optimize the entire expression for 
simplicity). Substitution of this value yields 
\begin{eqnarray*}
1 - \frac{\rho^{2}(\theta) }{ 1 + p/(N\theta)}\left( 1 - \frac{2}{N \theta}\right).
\end{eqnarray*}
Now let $\epsilon > 0$ be sufficiently small so that $\theta^{2} - \epsilon \theta(1 + \theta) > 0$, and let
$N$ be sufficiently large so that $\frac{2}{N \theta} < \epsilon$.
%\begin{eqnarray*}
%\frac{2}{N \theta} < \epsilon
%\end{eqnarray*}
We can always do this since $\theta$ only depends on $D$ and on the autocorrelation function.
%\leq 1 - \frac{1}{2}\frac{\rho^{2}(\theta) }{ 1 + p/(N\theta)}$
%as an upper bound in (\ref{eq:notdifficult}), where the latter is a consequence of the assumption 
%that $N \theta \geq 4$. 
Now suppose that $p/N = \theta^{2} - \epsilon \theta(1 + \theta)$, then 
\begin{eqnarray*}
1 - \frac{\rho^{2}(\theta) }{ 1 + p/(N\theta)}\left( 1 - \frac{2}{N \theta}\right) &\leq & 1 - \frac{\rho^{2}(\theta) }{ 1 + p/(N\theta)}( 1 - \epsilon) \\
& = & 1  - \frac{\rho^{2}(\theta)}{1 + \theta}  \leq D.
\end{eqnarray*}
The above implies that for $N$ sufficiently large, 
$\frac{1}{N}p_{\max}(N,D,\rho) \geq \theta^{2} - \epsilon \theta(1 + \theta)$.
%\begin{eqnarray*}
%\frac{1}{N}p_{\max}(N,D,\rho) \geq \theta^{2} - \epsilon \theta(1 + \theta).
%\end{eqnarray*}
Taking the liminf, we obtain that for all sufficiently small $\epsilon>0$,
\begin{eqnarray*}
\liminf_{N \rightarrow \infty} \frac{1}{N}p_{\max}(N,D,\rho) \geq \theta^{2} - \epsilon \theta(1 + \theta).
\end{eqnarray*}
%Taking the limit as $\epsilon \rightarrow 0$, we obtain the desired lemma conclusion. $\hfill\diamond$
Since $\epsilon > 0$ can be arbitrarily small, we obtain the desired conclusion. $\hfill\diamond$
\smallskip

The purpose of this Lemma is only to establish that $p_{\max}(N,D,\rho)$ grows at least linearly with $N$. 
The constants presented were chosen for simplicity of presentation.

The following is our main result on the rate of distributed coding: 
%\marginpar{\color{red}REQD? If allowing non-Gaussian sources, we do not need to restrict to a Gaussian
%$\bl{Z}$ any more in~\thmref{tungberg}. We must use the same $\bl{Z}$ here. Continued...} 
\begin{prop}
\proplbl{sumrate}
The sum rate of the distributed coding scheme described above is bounded above by 
a constant, independent of $N$. 
%$\frac{1}{2\theta^2}$, where $\theta$ is as in~\lemref{lingrowth}.
%Since the rate for each sensor is the same due to symmetry, the per-sensor rate therefore
%decreases inversely with the number $N$ of sensors for all large enough $N$. 
%decays as $\frac{1}{2N\theta^2}$ for all large enough $N$.
\end{prop}
{\em Proof:} Consider a vector Gaussian channel with input $\bl{W}\in\R^{N}$ and output $\bl{Y}\in\R^N$,
$\bl{Y} = \bl{W} + \bl{Z}$, where $\bl{Z}$ is as above,~and
where the power constraint on the input is given by $\cE[\bl{W}^T\bl{W}] \leq N$. 
Since $\bl{Z}$ is distributed $N(0,pI)$, the capacity of this
channel, 
\[
\max_{\bl{W}} I(\bl{W};\bl{W}+\bl{Z})~\mathrm{subject~to~} \cE[\bl{W}^T\bl{W}] \leq N,
\]
%\marginpar{\color{red}If it turns out that
%\thmref{tungberg} can be stated in the form that it is now in, with only second order properties
%involved, then we still choose Gaussian $\bl{Z}$ to get the channel with the least capacity.}
is equal to $\frac{N}{2}\log\left(1 + \frac{1}{p}\right)$ (see, for example,~\cite{tcbook}). 

Let $\epsilon > 0$ be any number smaller than $D_{net}$. 
We know from~\secref{eadsc} that there is an $N_1$ such that for $N \geq N_1$, 
$D'(N) \geq D_{net} -\epsilon$. Further, from~\lemref{lingrowth}, we know that there exists some 
$N_2 \geq 0$ and a constant 
%$\tilde{\theta} > 0$ 
$\theta > 0$ such that for
$N \geq N_2$, $p_{\max}(N,D_{net}-\epsilon,\rho) \geq {\theta}^2 N$. Clearly,
$p_{\max}(N,D,\rho)$ is a non-decreasing function of $D$, and therefore for $N\geq \max\{N_1, N_2\}$,
$p_{\max}(N,D'(N),\rho) \geq p_{\max}(N,D_{net}-\epsilon,\rho)$. It then follows that for 
$N \geq \max\{N_1, N_2\}$,
\[
I(\bl{X};\bl{X}+\bl{Z}) \leq \frac{N}{2}\log\left(1 + \frac{1}{{\theta}^2 N}\right).
\]
Then, using the inequality $\log(1+x) \leq x$, and using the result of~\thmref{tungberg} to 
substitute $\sum_{k=1}^{N} R_k$ for $I(\bl{X};\bl{X}+\bl{Z})$, we see that 
\[
\sum_{k=1}^{N} R_k = \frac{1}{2{\theta}^2}
\]
is achievable.
$\hfill\diamond$
\smallskip

The constants in~\propref{sumrate} have been chosen for simplicity. In general, the rates
achievable by distributed coding are smaller than the bound found in~\propref{sumrate}. 

\section{Comparison with a reference scheme}\seclbl{ref}

%\marginpar{\color{red}REQD: We are finding a {\em lower bound} on the rate of a scheme. But we know 
%that for centralized coding, Gaussian sources are the hardest to code. So, here the assumption
%for Gassianity is required. Otherwise, we would have to find a field that has the smallest rate
%distortion for a given MSE.}
In this section, we compare the rate of the distributed coding scheme discussed in~\secref{dsc}
with a reference scheme, which for reasons that will become apparent below, we call as
{\em centralized} coding.

The scheme consists of {\em one} centralized encoder $f$, which has access to samples taken 
at all sensors at times $\{1,\ldots,m\}$, 
and $N$ decoders, $\{g_k\}_{k=1}^{N}$ at the fusion center. The encoder maps 
the samples of the sensors, $X^{(1,\ldots,m)}(s_{1},\ldots,s_{N})$, into an index chosen
from the set $\{1,2,\ldots,\lfloor e^{m R^*_N}\rfloor\}$, where $R^*_N$ is the rate 
of the centralized scheme, and communicates this index to the fusion center. 
The decoder $g_k$ at the fusion center reconstructs the samples from sensor $k$ from the
messages received from the centralized encoder, 
\begin{eqnarray}
\nno \tilde{X}^{(1,\cdots,m)}(s_{k}) = g_{k}(f(X^{(1,\ldots,m)}(s_{1},\ldots,s_{N}))),
% \eqnlbl{jointencoding}
\end{eqnarray}
for $k=1,\ldots,N$.

At the fusion center, the reconstruction of the field $\tilde{X}(s)$ is obtained in the 
same two-step manner described in~\secref{dscrec}: the fusion center constructs estimates
$\tilde{X}(s_k)$ of the samples $X(s_k)$, for $k=1,\ldots,N$ from the messages received
from the sensors, and then interpolates between samples using~\eqnref{interpfn}. 

Let $R^*_N(D_{net})$ be the smallest rate for which there exists an encoder $f$ and 
decoders $\{g_k\}_{k=1}^N$ such that the integrated MSE~\eqnref{costfn} achieved by the above
scheme satisfies the constraint~\eqnref{errorcrit}. Then, it is clear that $R^*_N(D_{net})$
is a lower bound on the rates of all
schemes which use the two-step reconstruction procedure of~\secref{dscrec}. In this section we
bound the excess rate of the distributed coding scheme of~\secref{dsc} over the rate
$R^*_N(D_{net})$ of the centralized scheme. 

\subsection{Error analysis}\seclbl{eacomp}

Using the lower bound in~\appref{bounds}, equation~\eqnref{lbf}, on the error~\eqnref{costfn} in
terms of $J_{MSE}'(m)$ of~\eqnref{jprimemse} we conclude that for $N$ large enough,
if $J_{MSE}(m) \leq D_{net}$,  then $J_{MSE}'(m) \leq D''(N)$, where
%\marginpar{\color{red}NOT REQD:\\ This error analysis again uses only second order properties.}
\bqas
D''(N) = \frac{2\left(1-\rho^2\left(\frac{1}{2N}\right)\right) + 
2\sqrt{\left(1-\rho^2\left(\frac{1}{2N}\right)\right)
\left(1-\rho^2\left(\frac{1}{2N}\right) + D_{net}\right)} + D_{net}}{\rho^2\left(\frac{1}{2N}\right)}
\eqas
Note that $D''(N)$ approaches $D_{net}$ from above as $N\rightarrow\infty$.
%Note that for any $\epsilon > 0$, there is a $N_0$ large enough so that for $N > N_0$, $D''(N) \leq D_{net} + \epsilon$.

\subsection{Bounding the rate loss}
Now, consider
\bq
\eqnlbl{rdach}
\bl{V}^* = \arg\min_{p(\bl{V}|\bl{X})} I(\bl{X};\bl{V}),
\mathrm{subject~to~}\frac{1}{N}\cE\left[\|\bl{X}-\bl{V}\|^2_2\right] \leq D''(N).
\eq
From~\secref{eacomp}, it is clear that the rate of the centralized coding scheme, $R^*_N(D_{net})$ 
satisfies, for any $N$,
\[
R^*_N(D_{net}) \geq I(\bl{X};\bl{V}^*).
\]

%\marginpar{\color{red}We are again using the same $\bl{Z}$ as in~\thmref{tungberg}, and so we must recompute the
%capacity if the $\bl{Z}$ is not Gaussian.}
We now use techniques similar to those in~\cite{zamirloss} to bound the redundancy of 
distributed coding over the rate of joint coding. Let $\bl{Z}$ be as in~\propref{sumrate}. 
Expanding $I(\bl{X};\bl{X}+\bl{Z},\bl{V})$ in two ways, we get 
$I(\bl{X};\bl{X}+\bl{Z}) + I(\bl{X};\bl{V}|\bl{X}+\bl{Z})
= I(\bl{X};\bl{V})+ I(\bl{X};\bl{X}+\bl{Z}|\bl{V})$,
so that
\bqa
\eqnlbl{redpt} I(\bl{X};\bl{X}+\bl{Z})-I(\bl{X};\bl{V})&\leq&I(\bl{X};\bl{X}+\bl{Z}|\bl{V})\\
\nno&=&I(\left(\bl{X}-\bl{V}\right);\left(\bl{X}-\bl{V}\right)+\bl{Z}|\bl{V}).
\eqa
Since $\bl{V}\mch (\bl{X}-\bl{V}) \mch (\bl{X}-\bl{V})+\bl{Z}$, we have 
$I(\left(\bl{X}-\bl{V}\right);\left(\bl{X}-\bl{V}\right)+\bl{Z}|\bl{V}) 
\leq I(\left(\bl{X}-\bl{V}\right);\left(\bl{X}-\bl{V}\right)+\bl{Z})$.  
Subject to the constraint in~\eqnref{rdach}, 
$I(\left(\bl{X}-\bl{V}\right);\left(\bl{X}-\bl{V}\right)+\bl{Z})$ 
is upper bounded by the
capacity of a parallel Gaussian channel, with noise $\bl{Z}$ and input $\bl{W} = \bl{X}-\bl{V}$,
the power constraint on which is given by $\frac{1}{N}\cE[\|\bl{W}\|^2] \leq D''(N)$. The capacity of 
this channel is~\cite{tcbook} $C = \frac{N}{2}\log\left(1 + \frac{D''(N)}{p}\right)$, and 
therefore
%\[
%I(\left(\bl{X}-\bl{V}\right);\left(\bl{X}-\bl{V}\right)+\bl{Z}) \leq \frac{N}{2}\log\left(1 + \frac{D''(N)}{p}\right).
%\]
%Now, 
from~\eqnref{redpt} and the definition~\eqnref{rdach} of $\bl{V}$ as the 
rate-distortion achieving random vector, we get
\bqas
\nno I(\bl{X};\bl{X}+\bl{Z})-R_N^*(D_{net}) &\leq& \frac{N}{2}\log\left(1 + \frac{D''(N)}{p}\right).\\
&\leq& \frac{N}{2}\frac{D''(N)}{p},
\eqas
where the second inequality follows because $\log(1+x) \leq x$. 
From~\secref{eacomp}, we know that for any $\epsilon > 0$, there is a $N_1$ large enough so that 
for all $N \geq N_1$, $D''(N) \leq D_{net} + \epsilon$, and we can choose the variance
$p$ of the entries of $\bl{Z}$ to be at least $N\theta^2$, where ${\theta}$ is as 
in~\lemref{lingrowth}, while still ensuring that $\bl{X}+\bl{Z}$ meets the requirements on 
the auxiliary random variable $\bl{U}$ of~\thmref{tungberg}.
Therefore, substituting $\sum_{i=1}^{N} R_i$ for $I(\bl{X};\bl{X}+\bl{Z})$, and using~\lemref{lingrowth} 
and the result of~\secref{eacomp} we get that for any $\epsilon > 0$, 
there is an $N_1$ large enough so that for all $N \geq N_1$, 
\bqa
\eqnlbl{redbd}
\sum_{i=1}^{N} R_i - R_N^*(D_{net})    &\leq& \frac{D_{net} + \epsilon}{2{\theta}^2}.
\eqa

We conclude that the rate of the distributed coding scheme of~\secref{dsc} is no more than
a constant (independent of $N$) more than the rate of a centralized coding scheme with the 
same reconstruction procedure. 
Again, the constant in~\eqnref{redbd} has been chosen for 
simplicity of presentation and is in general much larger than the actual excess of the 
rate of the distributed coding scheme (see~\secref{numerical}).

\section{Point-to-point coding}\seclbl{p2p}

The distributed coding scheme studied in~\secref{dsc} shows that the tradeoff of sensor
numbers to sensor accuracy is achievable. However, it may not be feasible to implement
complicated distributed coding schemes in simple sensors. In this section we show that 
if the sensors are synchronized and if a delay that increases linearly with the number 
of sensors is tolerable, then the desired tradeoff can be achieved by a simple scheme in 
which encoding can be performed at sensors without any knowledge of the correlation 
structure of the field. 

In this scheme, we partition the interval $[0,1]$ into $K$ equal sized sub-intervals, 
$[0,\frac{1}{K}],(\frac{1}{K},\frac{2}{K}]$,$\ldots$,$(\frac{K-1}{K},1]$. We specify $K$ later, but assume that
$N>K$ sensors are placed uniformly in $[0,1]$. We assume that $K$ divides $N$ for simplicity (so that
there are an integer number, $\frac{N}{K}$, of samples in each interval). 

Since the somewhat involved 
notation may obscure the simple idea behind the scheme, we explain it before describing the scheme in detail. 
We consider time in blocks of duration $\frac{N}{K}$ units each. The scheme
operates overall with a blocklength of $m = m'\frac{N}{K}$, that is, $m'$ blocks, for some integer $m'$. 
%At any time, exactly one sensor from each sub-interval is active, and 
Each sensor %in a sub-interval
is active exactly once in any time interval that is $\frac{N}{K}$ units in duration. 
A sensor samples the field at its location only at those times when it is active. 
Each sensor uses a point-to-point code of blocklength $m'$ and rate $R_p$ nats per {\em active} 
time unit. The code is chosen appropriately so as to meet the distortion constraint. However, 
since the sensor is active only in $m'$ out of $m'\frac{N}{K}$ time units, the rate of the 
code {\em per time-step} is only
$\frac{K}{N}R_p$ nats. We show below that the desired distortion can be achieved with a rate $R_p$ 
that is independent of $N$ and therefore the desired scaling can be achieved by the above scheme.

We now describe the scheme in detail. Consider the time instants $\left\{1,2,\ldots,m'\frac{N}{K}\right\}$. 
Each sensor uses a code of blocklength $m = m'\frac{N}{K}$, which is constructed from
a code of blocklength $m'$, as follows. For each $j$ in $\{1,2,\ldots,\frac{N}{K}\}$ and each
$l$ in $\{0,1,\ldots,K-1\}$, sensor ${\frac{N}{K}l+j}$ (which is the $j$-th sensor from 
the left in the sub-interval $\left(\frac{l}{K},\frac{l+1}{K}\right]$, and is at 
location $s_{\frac{N}{K}l+j}$) samples the field only at times 
$\cT_{l,j} = \{j,j+\frac{N}{K},j+\frac{2N}{K},\ldots,j+\frac{(m'-1)N}{K}\}$. It uses a code 
of rate $R_p$, to be specified below, to map the $m'$ samples $\{X^{(i)}(s_{\frac{N}{K}l+j}), i \in \cT_{l,j}\}$ to an 
element of the set $\{1,2,\ldots,\lfloor e^{m'R_p}\rfloor\}$. The rate per-time unit of each 
sensor is therefore $\frac{1}{m'\frac{N}{K}}m'R_p = \frac{K}{N}R_p$ nats.

The fusion center consists of $N$ decoders, one for each sensor. Decoder $k$ constructs 
estimates of the samples encoded by sensor $k$ {\em using only messages received from
sensor $k$}. Then, for each time 
$i = \frac{N}{K}l + j$ in $\{1,\ldots,m'\frac{N}{K}\}$, the fusion center has reconstructions 
\[
\left[\tilde{X}^{(i)}(s_j), \tilde{X}^{(i)}(s_{\frac{N}{K}+j}), \tilde{X}^{(i)}(s_{\frac{2N}{K}+j}), 
\ldots, \tilde{X}^{(i)}(s_{\frac{(K-1)N}{K}+j})\right],\] that is, one reconstruction for each sub-interval.

For any $s\in[0,1]$, we denote the location of the (unique) sensor active within the 
interval $(\frac{l}{K},\frac{l+1}{K}]$ to which $s$ belongs by $r^{(i)}(s)$. For each time
instant $i$, the fusion center reconstructs the field for $s\neq r^{(i)}(s)$ as 
\[
\tilde{X}^{(i)}(s) = \rho(s-r^{(i)}(s))\tilde{X}^{(i)}(r^{(i)}(s)),
\]
where $\tilde{X}^{(i)}(r^{(i)}(s))$ is the decoded sample at the fusion center of the sensor 
at $r^{(i)}(s)$ at time $i$. 

%\marginpar{\color{red}CAN DO: This error analysis {\em does} use the assumption of Gaussianity. However, 
%we can do a coarser analysis as in~\appref{bounds} to make it depend only on second order
%properties.}
We show in~\appref{eap2p} that
\bqa
\nno \lefteqn{\frac{1}{m}\sum_{i=1}^{m} \int_0^1 \cE[(X^{(i)}(s)-\tilde{X}^{(i)}(s))^2] ds} \\
\eqnlbl{p2pbd} &\leq&(1-\rho^2(\frac{1}{K})) + 
\frac{1}{N}\sum_{k=1}^{N} 
\left\{\frac{1}{m'} \sum_{i_k\in\cT_k}\cE[\left(X^{(i_k)}(s_k)-\tilde{X}^{(i_k)}(s_k)\right)^2]\right\}
\eqa
where, with some abuse of notation, we use $\cT_k$ to denote the set of time steps in which 
sensor $k$ is active. Note that the cardinality of $\cT_k$ is $m'$ for each $k$. 

We now choose $K$ large enough so that $(1-\rho^2(\frac{1}{K})) < D_{net}$ and choose 
\bq
\eqnlbl{p2pd}
D_K = D_{net} - (1-\rho^2(\frac{1}{K})).
\eq
%\marginpar{\color{red}NOT REQD\\
%Here we are upper bounding the rate, and so using Gaussian sources (hardest to code) is acceptable.}
The $m'$-blocklength code used at sensor $k$ for the times that it is active is a code that
achieves the rate-distortion bound for the distortion constraint
\[
\frac{1}{m'} \sum_{i_k\in\cT_k}\cE[\left(X^{(i_k)}(s_k)-\tilde{X}^{(i_k)}(s_k)\right)^2 \leq D_K,
\]
as $m'\rightarrow\infty$. It is
well known that the rate of this code is $R_p = \frac{1}{2}\log\frac{1}{D_K}$ nats per time step. 
It is clear from~\eqnref{p2pbd} and~\eqnref{p2pd} that this scheme achieves the required distortion. 
Since the rate of each sensor in the overall scheme is $\frac{K}{N}R_p$ 
nats per time step we have therefore constructed a scheme  
in which the bit rate of each sensor is
\bq
\eqnlbl{p2prate}
-\frac{K}{N}\frac{1}{2}\log\left[D_{net} - (1-\rho^2(\frac{1}{K}))\right]
\eq
nats per time step. We can now choose $K$ to minimize the sum-rate 
$-\frac{K}{2}\log\left[D_{net} - (1-\rho^2(\frac{1}{K}))\right]$. 

Further, it is well known (see~\cite[Section 5.1]{toby}) that using scalar quantization,
each sensor can achieve distortion $D_K$ at rate $\frac{1}{2}\log\frac{1}{D_K} + \delta$, 
where $\delta$ is a small constant. For example, for Max-Lloyd quantizers 
(see~\cite[Section 5.1]{toby}), $\delta$ is less than $1$ bit. 

Therefore, we conclude that it is indeed possible to achieve the desired tradeoff between
sensor numbers and the per-sensor rate even when the sensors encode their measurements using
appropriate scalar quantizers, given that we also make use of the synchronization between sensors
to activate sensors appropriately. This is in contrast to the conclusions of~\cite{neuhoff},
where full use of synchronization is not made, and therefore it is found that the above tradeoff
is not achievable with scalar quantization. 

\section{Numerical examples}\seclbl{numerical}

In this section we give numerical examples of the rates of the coding schemes discussed
in~\secref{dsc},~\secref{ref} and~\secref{p2p}. The two fields we consider as
examples are (1) a (spatially) band-limited Gaussian field, for which $\rho(\tau) = \sinc(\tau)$, 
where $\sinc(\tau)=\frac{\sin(\pi\tau)}{\pi\tau}$, and (2) a Gauss-Markov field, for which
$\rho(\tau) = \exp\{-|\tau|\}$. 

For these fields, we numerically find the largest value $p_{\max}$ of the variance $p$ of $\bl{Z}$ for
which the error for the estimator in~\eqnref{optest} is no more than the distortion $D'(N)$ 
of~\eqnref{dprime}, with $D_{net}  =0.1$. The resulting values are shown in~\figref{pdvar}. We see
that for large values of $N$, $p_{\max}$ is indeed approximately linear in $N$. 
\begin{figure}
\centerline{\includegraphics*[width=6cm]{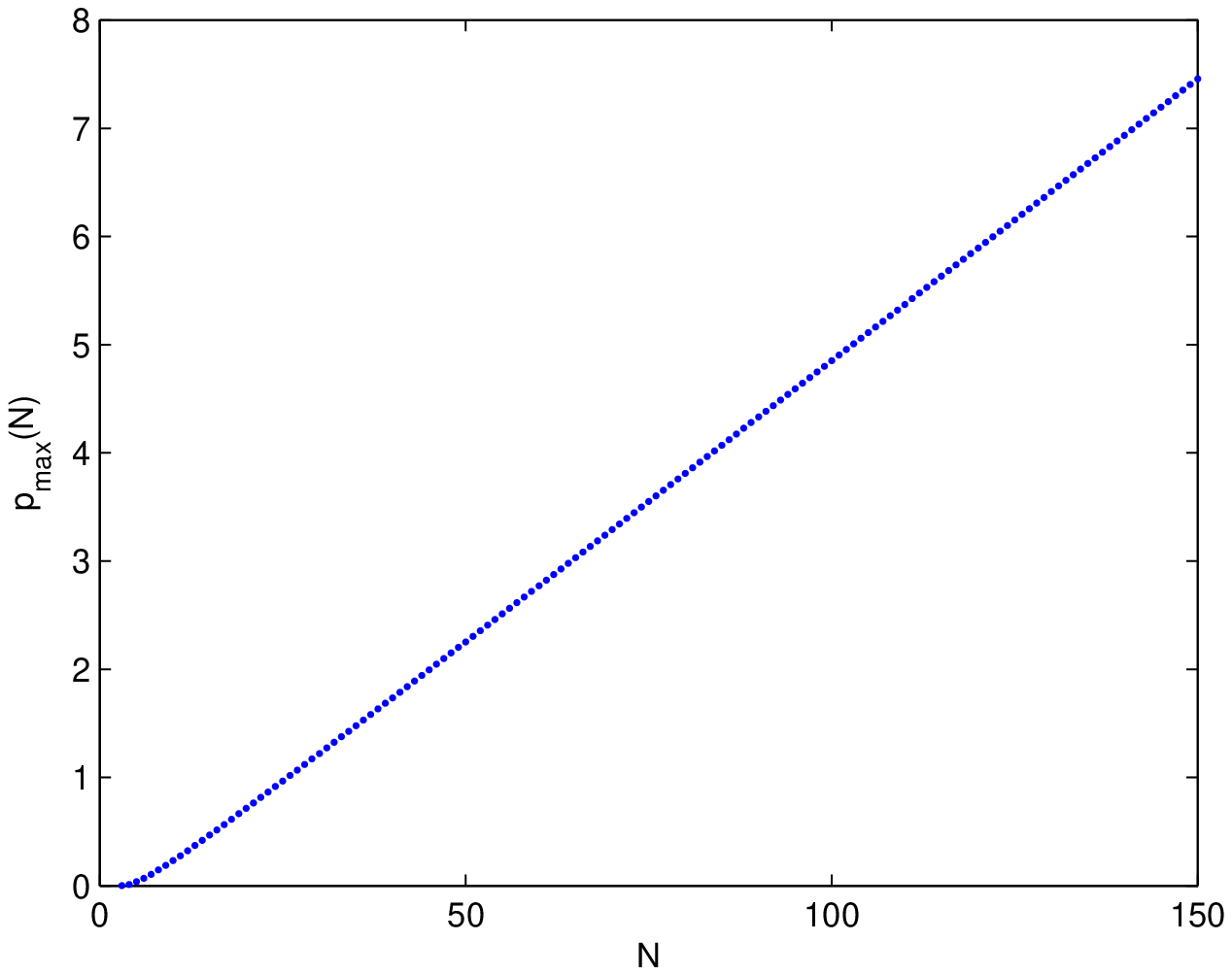}
\includegraphics*[width=6cm]{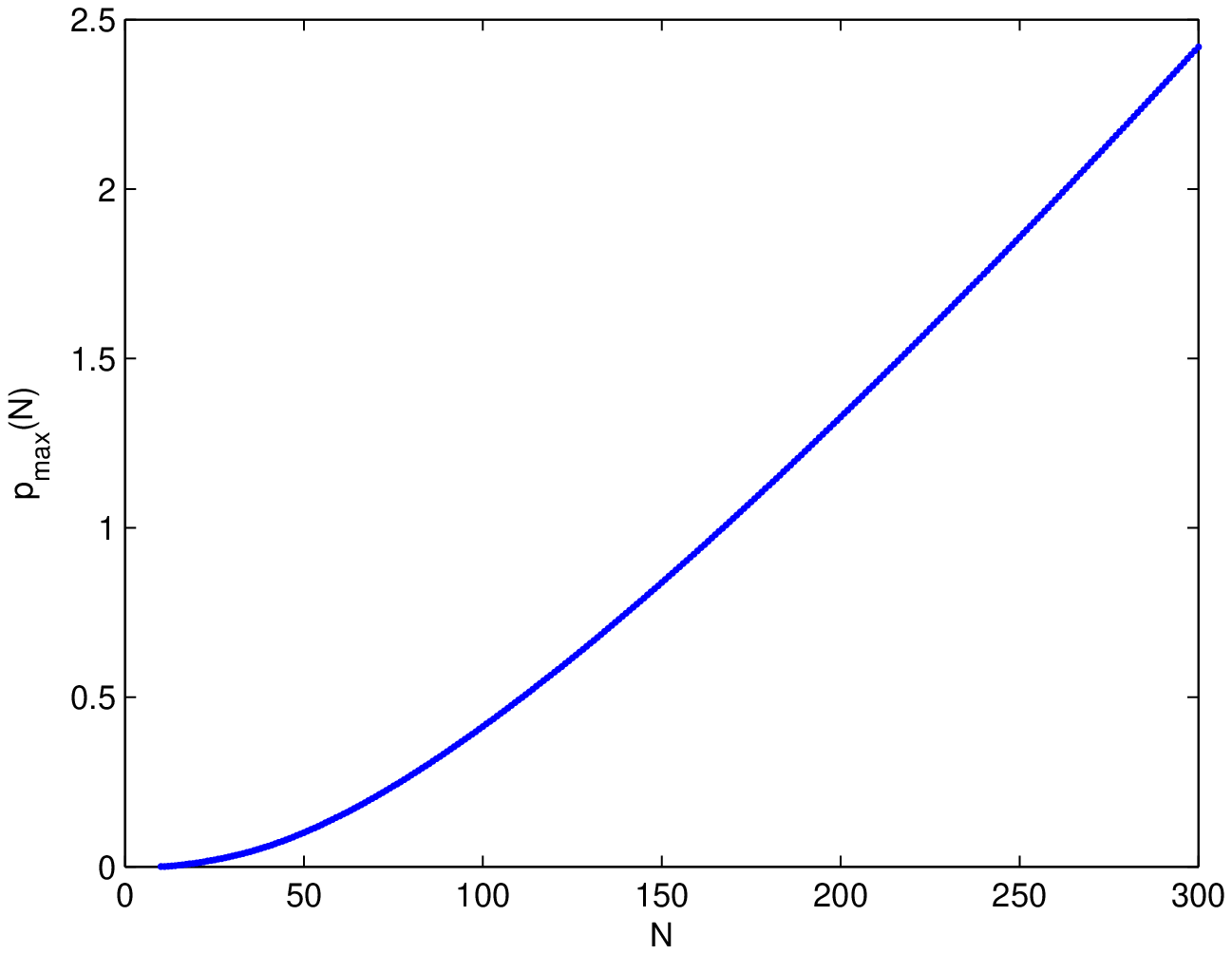}}
\caption{\figlbl{pdvar}Linear increase of $p_{max}$ for large $N$: 
$\rho(\tau) = \sinc(\tau)$ (left) and $\rho(\tau) = \exp\{-|\tau|\}$ (right). $D_{net} =0.1$.}
\end{figure}

We compute the achievable sum rate of the distributed source coding scheme, which is equal 
to $I(\bl{X};\bl{X}+\bl{Z})$ from~\thmref{tungberg}, with the $p_{\max}$ found above as 
the variance of the entries of $\bl{Z}$. These rates are shown in~\figref{sumrates}. 
For reference, we also show the lower bound on the 
rate of the centralized coding scheme computed in~\secref{ref}. 
\begin{figure}
\centerline{\includegraphics*[width=7cm]{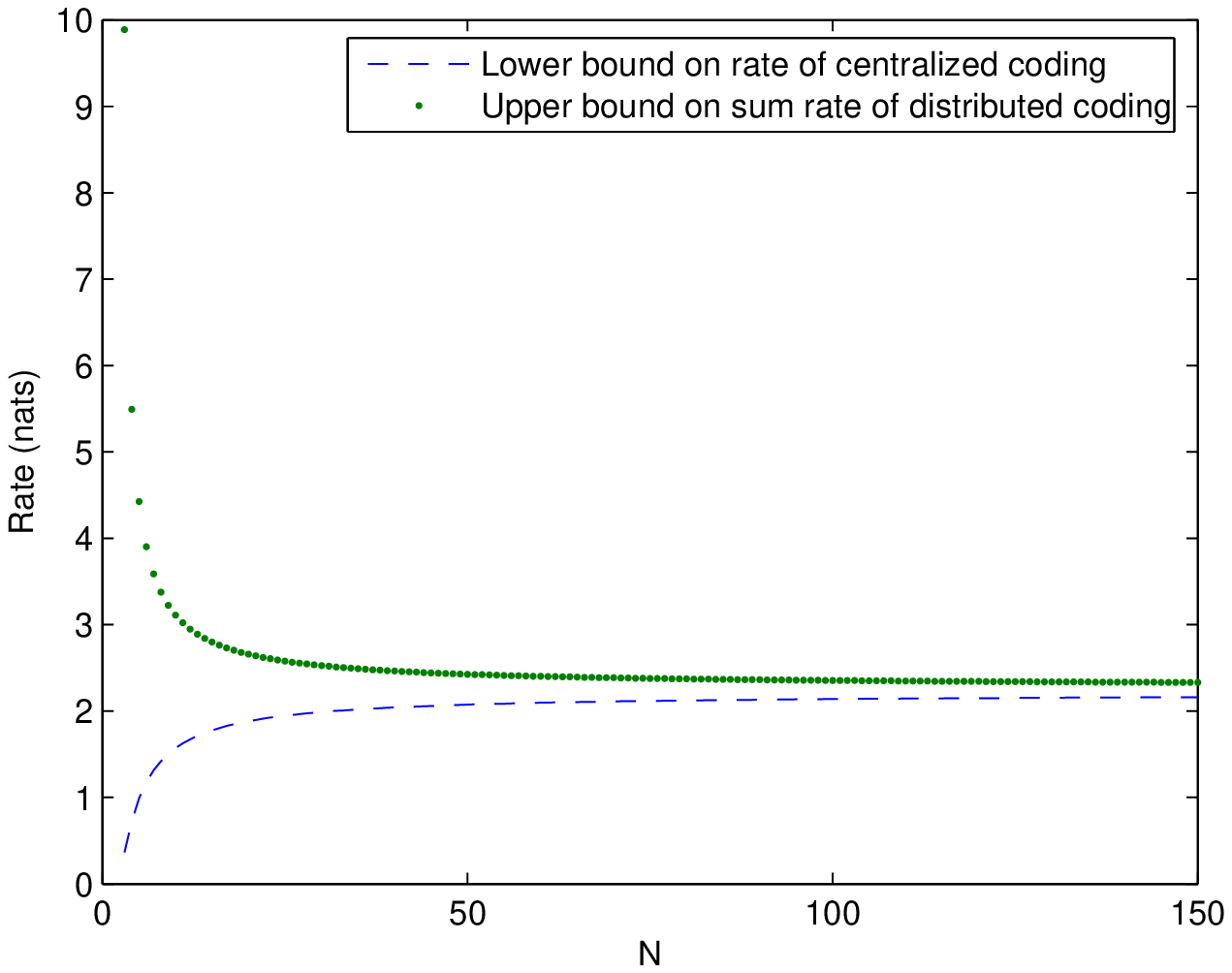}
\includegraphics*[width=7cm]{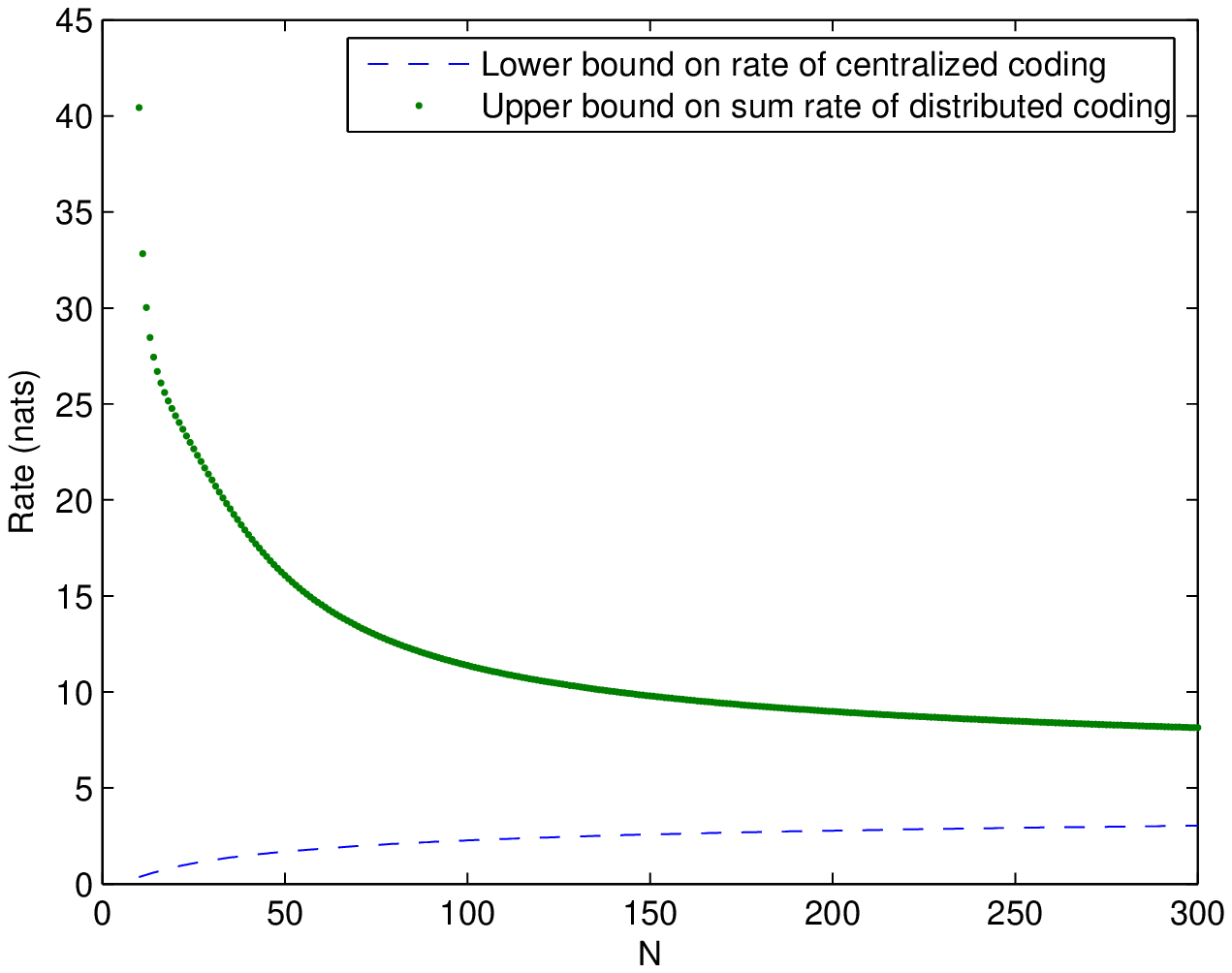}}
\caption{\figlbl{sumrates}Rates of joint and distributed coding (in nats per snapshot) vs.
number of sensors $N$: $\rho(\tau) = \sinc(\tau)$ (left) and $\rho(\tau) = \exp\{-|\tau|\}$ 
(right). $D_{net} =0.1$.}
\end{figure}

In comparison, on minimizing the rate~\eqnref{p2prate} of the point-to-point coding 
scheme of~\secref{p2p}, we find that best sum rate for $\rho(\tau) = \sinc(\tau)$ is 
$11.77$ nats for $K=7$ intervals, and that the best sum rate for $\rho(\tau) = \exp(-|\tau|)$ 
is $46.92$ nats with $K=24$ intervals, which is significantly greater than the sum-rate of 
the distributed coding scheme found above. However, part of the reason
for the large sum-rate of the point-to-point coding scheme is that our analysis exaggerates
an edge-effect for the sake of simplicity:
In~\secref{p2p} we estimated the value of the field at point $s$ at time $i$ using 
the sample that the fusion center has at time $i$ from the sub-interval that $s$ lies in. 
We could instead have used the sample {\em closest} to $s$ that is available at the fusion 
center at time $i$, similar to what is done in~\secref{dsc} and~\secref{ref}. However, this 
would have meant dealing with the first and the last sub-interval differently, and therefore
we did not follow the analysis outlined above. Without this edge effect, the rates of the 
point-to-point coding scheme are approximately half the rates found above, which are still
considerably larger than the sum-rates of the distributed coding scheme.

\section{Conclusions}\seclbl{conclusion}

We have studied the sum rate of distributed coding for the reconstruction of
a random field using a dense sensor network. We have shown the existence of a distributed
coding scheme which achieves a sum rate that is a constant independent of the number of
sensors. Such a scheme is interesting because it allows us to achieve a per-sensor rate
that decreases inversely as the number of sensors, and therefore to achieve small per-sensor
rates using a large number of sensors. 

In obtaining bounds on the sum rate of distributed coding, we made full use to the 
heavy correlation between samples of the field taken at positions that are close 
together. When the 
number of sensors is large, the redundancy in their data can be utilized by coding more
and more coarsely: this corresponds to more noisy samples, and is manifested in the 
growth of the noise $p_{\max}$ in the forward channel in~\secref{dsc}. We believe that
this technique of bounding the sum rate is of independent interest.

We have also shown that contrary to what has been suggested in~\cite{neuhoff} and~\cite{neupra}, 
it is indeed possible to design a scheme that achieves a constant sum rate with sensors that
are scalar quantizers, even {\em without} the use of distributed coding. This scheme, however, requires
that we make appropriate use of the synchronization between the sensors, results in 
a delay in reconstruction which increases linearly with the number of sensors, and achieves
rates that may be significantly higher than the rates achieved by distributed coding. The scheme
is nevertheless interesting because its low complexity makes it easy to implement. 

\section*{Acknowledgement}
The first author thanks Prof. R. Srikant for many insightful comments on this work,  
and for his encouragement to work on this paper while the first author was at UIUC. 

\appendix
%\marginpar{\color{red}NOT REQD\\This error analysis uses only second order properties of the field, 
%and therefore remains unchanged upon relaxing the assumption of Gaussianity.}
\section{Bounds on $J_{MSE}(m)$ for the schemes in~\secref{dsc} and~\secref{ref}}\applbl{bounds}

We can write the error in reconstruction at any $s\in [0,1]$ as 
\bqa
\nno X(s) - \tilde{X}(s) &=& X(s) - \rho(s-n(s))\tilde{X}(n(s))\\
\nno &=& \left[ X(s) - \rho(s - n(s)) X(n(s)) \right] 
+ \left[ \rho(s - n(s))\left(X(n(s)) - \tilde{X}(n(s))\right) \right]\\
\eqnlbl{ersplit}
&=& E_{S}(s) + E_{Q}(s),
\eqa
where $E_{S}(s) =  X(s) - \rho(s - n(s)) X(n(s))$ and 
$E_Q(s) = \rho(s - n(s))\left(X(n(s)) - \tilde{X}(n(s))\right)$. 
Note that in the schemes described in~\secref{dsc} and~\secref{ref}, the encodings 
of all samples are used to obtain the estimate $\tilde{X}(n(s))$, and therefore
$\tilde{X}(n(s))$ is in general not independent of $X(s_k)$, for $s_k\neq n(s)$. As 
a result, $E_S(s)$ and $E_Q(s)$ are in general not independent. In this appendix, 
we find upper and lower bounds on $J_{MSE}(m)$ that hold for the schemes of~\secref{dsc}
and~\secref{ref}.

Using the Cauchy-Schwarz inequality (for any two appropriately integrable random variables
$A$ and $B$, $|\cE[AB]| \leq \sqrt{\cE[A^2]\cE[B^2]}$), it is easy to see that 
\bqa
\eqnlbl{ub}
\cE\left(E_S(s) + E_Q(s)\right)^2 
&\leq&
\cE\left(E_S(s)\right)^2 + \cE\left(E_Q(s)\right)^2
+ 2\sqrt{\cE\left(E_S(s)\right)^2\cE\left(E_Q(s)\right)^2}\\
\eqnlbl{lb}
\cE\left(E_S(s) + E_Q(s)\right)^2 
&\geq&
\cE\left(E_Q(s)\right)^2
- 2\sqrt{\cE\left(E_S(s)\right)^2\cE\left(E_Q(s)\right)^2}.
\eqa
Now, note that $\cE\left(E_S(s)\right)^2 = (1-\rho^2(s-n(s))$. Therefore, 
\bqa
\nno \cE\left(E_S(s)\right)^2\cE\left(E_Q(s)\right)^2 &=& 
\rho^2(s-n(s))\left(1-\rho^2(s-n(s))\right)\cE\left(X(n(s))-\tilde{X}(n(s))\right)^2.
\eqa
For $N$ large enough so that both $\rho^2\left(\frac{1}{2N}\right) \geq \frac{1}{2}$ and $1/(2N)$ 
lies in the interval around 
$0$ in which $\rho$ is non-increasing (so that for $s\in\left(\frac{k}{N},\frac{k+1}{N}\right)$ 
$\rho^2(s-n(s))(1-\rho^2(s-n(s)) \leq\rho^2(\frac{1}{2N})(1-\rho^2(\frac{1}{2N}))$,
which holds because the function $h(x)=x(1-x)$ is decreasing in $[\frac{1}{2}, 1]$), we get 
that
\bqa
\eqnlbl{sqrtbd}
\cE\left(E_S(s)\right)^2\cE\left(E_Q(s)\right)^2 &\leq&
\rho^2\left(\frac{1}{2N}\right)\left(1-\rho^2\left(\frac{1}{2N}\right)\right)\cE\left(X(n(s))-\tilde{X}(n(s))\right)^2.
\eqa

From~\eqnref{costfn} and~\eqnref{ersplit}, we have 
\bq
\eqnlbl{ersplitav}
J_{MSE}(m) = \frac{1}{m}\sum_{i=1}^m
\int_0^1 \cE\left(E_{S}^{(i)}(s) + E_{Q}^{(i)}(s) \right)^2ds.
\eq
Therefore, integrating~\eqnref{ub} and~\eqnref{lb} over $[0,1]$, using~\eqnref{sqrtbd} and 
Jensen's inequality (and the concavity of the function $y(x) = \sqrt{x}$), and averaging over
the time index, we get 
\bqa
\eqnlbl{ubf}
J_{MSE}(m) &\leq& \left\{1-\rho^2\left(\frac{1}{2N}\right)\right\} + J_{MSE}'(m) + 
2\sqrt{\rho^2(\frac{1}{2N})(1-\rho^2(\frac{1}{2N}))J_{MSE}'(m)},\\
\eqnlbl{lbf}
J_{MSE}(m) &\geq& \rho^2(\frac{1}{2N})J_{MSE}'(m) - 
2\sqrt{\rho^2(\frac{1}{2N})(1-\rho^2(\frac{1}{2N}))J_{MSE}'(m)},
\eqa
where $J_{MSE}'(m)$ is as in~\eqnref{jprimemse}.

%\marginpar{\color{red}REQD/CAN DO: Here we have made use of Gaussianity, but we can relax that
%by doing a coarser analysis as in~\appref{bounds} above.}
\section{Error analysis for the point-to-point coding scheme}\applbl{eap2p}

With some abuse of notation, we can still write the error in reconstruction as
\[
X(s) - \tilde{X}(s) =  E_S(s) + E_Q(s),
\]
where now 
\bqa
\nno E_S(s) &=& X(s) - \rho(s-r(s))X(r(s)), ~\mathrm{and}\\
\nno E_Q(s) &=& \rho(s-r(s))\left(X(r(s))-\tilde{X}(r(s))\right).
\eqa
In the point-to-point coding scheme, the fusion center estimates the samples of each sensor
using only the messages that it receives from that particular sensor. 
Note that $E^{(i)}_S(s)$ is the error in the optimal MSE estimate of $X(s)$ given $X^{(i)}(r(s))$. 
It is well known that if $\{X(s),s\in [0,1]\}$ is a Gaussian process, the error $E^{(i)}_S(s)$ in is 
independent of $X^{(i)}(r^{(i)}(s))$. 
Further, due to the independence of the field $X^{(i)}$ and 
the field $X^{(j)}$ for any $j\neq i$, $E^{(i)}_S(s)$ is independent
of $X^{(j)}(r^{(j)}(s))$ for all $j$, and hence also of the reconstructions 
$\tilde{X}^{(j)}(r^{(j)}(s))$ and the error terms $E^{(i)}_Q(s)$. Therefore,
for any $i$, 
\[
\cE[(X^{(i)}(s)-\tilde{X}^{(i)}(s))^2] = \cE[(E_S^{(i)}(s))^2] + \cE[(E_Q^{(i)}(s))^2].
\]
Now, for $K$ large enough, $\cE[(E_S^{(i)}(s))^2]  = 1-\rho^2(s-r^{(i)}(s)) \leq 1-\rho^2(\frac{1}{K})$ 
for every $s\in [0,1]$. Also, since $\rho^2(s) \leq 1$ for all $s \in [0,1]$,
\bqa
\nno \cE[(E_Q^{(i)}(s))^2] % &=& \rho^2(s-r^{(i)}(s))\cE[\left(X^{(i)}(r^{(i)}(s))-\tilde{X}^{(i)}(r^{(i)}(s))\right)^2] \\
\nno &\leq& \cE[\left(X^{(i)}(r^{(i)}(s))-\tilde{X}^{(i)}(r^{(i)}(s))\right)^2].
\eqa
So, we get 
\bqa
\nno \int_0^1 \cE[(X^{(i)}(s)-\tilde{X}^{(i)}(s))^2]ds  &=& \sum_{l=0}^{K-1}\int_{\frac{l}{K}}^{\frac{l+1}{K}}\cE[(X^{(i)}(s)-\tilde{X}^{(i)}(s))^2]ds\\
\nno &\leq& \sum_{l=0}^{K-1}\int_{\frac{l}{K}}^{\frac{l+1}{K}}(1-\rho^2(\frac{1}{K})) + \cE[\left(X^{(i)}(r^{(i)}(s))-\tilde{X}^{(i)}(r^{(i)}(s))\right)^2]ds\\
\nno &=& (1-\rho^2(\frac{1}{K})) + \frac{1}{K}\sum_{l=0}^{K-1}\cE[\left(X^{(i)}(r^{(i)}(\frac{l+1}{K}))-\tilde{X}^{(i)}(r^{(i)}(\frac{l+1}{K}))\right)^2],
\eqa
where we note that by our notation, $r^{(i)}(\frac{l+1}{K})$ is the location of the (unique) sensor active at
time step $i$ in the interval $(\frac{l}{K},\frac{l+1}{K}]$.

Now summing over the time index we get,
\bqa
\nno \lefteqn{\frac{1}{m}\sum_{i=1}^{m} \int_0^1 \cE[(X^{(i)}(s)-\tilde{X}^{(i)}(s))^2]ds}\\
\nno &\leq&  (1-\rho^2(\frac{1}{K})) 
+ \frac{1}{Km}\sum_{i=1}^{m}\sum_{l=0}^{K-1}\cE[\left(X(r^{(i)}(\frac{l+1}{K}))-\tilde{X}(r^{(i)}(\frac{l+1}{K}))\right)^2].
\eqa
Rearranging the sum on the right and substituting $m = \frac{m'N}{K}$ we get
\bqas
\lefteqn{\frac{1}{m}\sum_{i=1}^{m} \int_0^1 \cE[(X^{(i)}(s)-\tilde{X}^{(i)}(s))^2]ds}\\
&\leq& (1-\rho^2(\frac{1}{K})) 
+ \frac{1}{m'N}\sum_{k=1}^{N}\sum_{i_k\in\cT_k}\cE[\left(X^{(i_k)}(s_k)-\tilde{X}^{(i_k)}(s_k)\right)^2],\\
&=& (1-\rho^2(\frac{1}{K})) 
+ \frac{1}{N}\sum_{k=1}^{N} 
\left\{\frac{1}{m'} \sum_{i_k\in\cT_k}\cE[\left(X^{(i_k)}(s_k)-\tilde{X}^{(i_k)}(s_k)\right)^2]\right\}
\eqas
where $\cT_k$ is the set of time steps in which sensor $k$ is active.

\bibliographystyle{IEEEtran}

\end{document}